# Assessing the effects of seasonal tariff-rate quotas on vegetable prices in Switzerland

Daria Loginova*, Marco Portmann**, Martin Huber+

November 2020

**Abstract:** Causal estimation of the short-term effects of tariff-rate quotas (TRQs) on vegetable producer prices is hampered by the large variety and different growing seasons of vegetables and is therefore rarely performed. We quantify the effects of Swiss seasonal TRQs on domestic producer prices of a variety of vegetables based on a difference-in-differences estimation using a novel dataset of weekly producer prices for Switzerland and neighbouring countries. We find that TRQs increase prices of most vegetables by more than 20% above the prices in neighbouring countries during the main harvest time for most vegetables and even more than 50% for some vegetables. The effects are stronger for more perishable vegetables and for conventionally produced ones compared with organic vegetables. However, we do not find clear-cut effects of TRQs on the week-to-week price volatility of vegetables although the overall lower price volatility in Switzerland compared with neighbouring countries might be a result of the TRQ system in place.

**Keywords:** agricultural trade policies, price volatility, Swiss agriculture, seasonal production, vegetable production, price stability, stabilisation policies, tariff-rate quotas, difference-in-differences, inverse probability weighting

**Adresses for correspondence:** *Daria Loginova, Agroscope, Tänikon 1, CH-8356 Ettenhausen, Tel +41 58 469 91 47, daria.loginova@agroscope.admin.ch, www.agroscope.admin.ch; **Marco Portmann, Economic Analysis Unit, Swiss Federal Tax Administration, Eigerstrasse 65, CH-3003 Bern, Tel +41 58 481 70 54, marco.portmann@estv.admin.ch, www.estv.admin.ch; +Martin Huber, Dept. of Economics, University of Fribourg, Bd de Pérolles 90, CH-1700 Fribourg, Tel +41 26 300 82 74, martin.huber@unifr.ch, www.unifr.ch/appecon/en/team/martin-huber/.

# I. INTRODUCTION

Support for agricultural producers is often achieved by trade barriers, such as tariff-rate quotas, or TRQs (Aksoy and Beghin, 2005; WTO, 2016).[1] These policies contribute to the gap between domestic and international prices, which is regularly analysed with several annual market price support estimates for selected product groups.[2] By contrast, the short-term effects of trade barriers on disaggregated vegetable prices are rarely identified causally. Exceptions are, for instance, Santeramo and Cioffi (2012), Márquez-Ramos and Martínez-Gómez (2016) and Hillen (2019), who examined only six different products. There are several reasons for the low number of studies. First, a large heterogeneity exists between vegetables in terms of perishability and biological characteristics in general, market structures and production techniques, applied policies and the level of protection. Second, seasonality in production and demand, short-term weather conditions and long-term climatic conditions as well as consumer preferences towards home-grown, domestic and imported vegetables can have geographically distinct ramifications that hamper international comparisons. Finally, price differentials between domestic and international prices might reflect quality differentiation. As Abbott (2012) pointed out, in short-term markets, even successful price support measures can cause short-term price volatility. Therefore, the analysis of the effect of trade regulations on price levels as well as the price volatility of fruits and vegetables deserve more attention.

In this article, we study the effects of Switzerland's comprehensive system of seasonal TRQs for vegetables on producer prices for the time horizon from 2014 to 2019. We draw on a unique dataset of weekly producer prices from Switzerland and from neighbouring regions in Italy, France and Germany, and we employ a difference-in-differences approach based on an inverse probability weighting estimator (see Abadie, 2005) to identify the effects of seasonal TRQs on producer price levels and stability in Switzerland. The setting is remarkable for several reasons. While out-of-season tariff rates for most vegetables are low, high tariff rates are imposed during the so-called 'administrated' or 'protected' period, which essentially covers the main harvest period and serves to protect domestic producers from foreign competitors. The start and end of this period are precisely fixed by laws and ordinances which define the period of treatment, that is, protection. Since we have weekly data and the prices of perishable goods react quickly, we can identify the short-term effects of TRQs on Swiss prices. As our analysis quantifies the effects relative to neighbouring countries, which are all EU member states, it is important to note that the EU imposes tariffs on some vegetable groups, too (see Appendix 1). While these EU tariffs are less stringent than the Swiss ones, they nevertheless imply that our estimates for these vegetables constitute a lower bound on the effect of a hypothetical comparison of Swiss TRQs vs. no TRQs at all. While many studies have focused on major crops, such as maize and wheat, we analyse 35 different kinds of vegetables. All vegetables have individually tailored protection periods and import quotas. Assessing the effects of TRQs on each vegetable in turn allows us to examine the effect of heterogeneity, for example, caused by differences in perishability. Finally, the estimated market price support is based on a comparison of domestic prices in the unprotected and the protected periods and a comparison of domestic prices to those in neighbouring countries. The comparison with neighbouring countries is necessary as a pure comparison of domestic prices cannot account for seasonality or other metrological effects. By using price data from Switzerland's neighbouring regions, we control for common influences. However, the chosen difference-in-differences approach does not hinge on similar price levels between the countries but on the common trend assumption to be satisfied (see e.g. Lechner, 2011). The common trend assumption requires that changes in producer prices over time in countries with and without TRQs would, on average, be the same if neither country imposed TRQs. This allows for differences in price levels across countries, caused by unobserved characteristics, as long as the price effect of the characteristics remains stable over time.

---

[1] Almost 60% of the total support for agricultural producers in the OECD, EU and key emerging economies was provided by keeping producer prices on domestic markets above international prices (OECD, 2017).
[2] OECD's market price support estimates are published either at the aggregate agricultural level or at the level of agricultural product groups.

Tariffs and import quotas can be set by authorities in consultation with market participants during the protected period (see the detailed explanation in Section II). Given the producers' involvement in the process and the intrinsic aim of market price support, our first hypothesis is that we expect Swiss producer prices to increase from the unprotected to the protected period and compared with neighbouring countries. Our results quantify the price support. With regard to short-term price stability, our expectations are less clear. The Swiss TRQ system allows virtually blocking imports on short notice if the domestic harvest outgrows the domestic demand. At the same time, authorities aim to avoid consumer price spikes by opening quotas in cases where the domestic harvest is temporarily short. Based on these arguments, one would expect lower price fluctuations. However, as the Swiss vegetable production takes place in a rather small geographic area with common metrological shocks, the protected periods create incentives to concentrate harvests within those periods. Such a concentration of the production could entail a reduction in short-term price stability.

Our results show that seasonal TRQs increase prices by more than 20% above the prices in neighbouring countries in the main harvest time for most vegetables and even more than 50% for some vegetables. The support is stronger for conventionally than for organically produced vegetables. The more perishable the vegetable is, the more it profits from support. We find only modest effects from seasonal TRQs on weekly price volatility.

The remainder of this article is structured as follows. Section II presents the relevant literature on TRQs. Section III describes the institutional background, and Section IV provides our data and empirical approach. Section V presents our results, and Section VI concludes the study.

## II. LITERATURE

### *TRQs and agricultural trade*

There was international consensus in the 1990s that trade barriers consisting of a complex and costly mix of tariff- and non-tariff measures should be simplified by tariffication, as decided by the Uruguay Round of the General Agreement on Tariffs and Trade (GATT, 1994).[3] Moschini (1991), and later others, including Abbott and Paarlberg (1998), questioned the success of the tariffication initiative. Eventually, by setting high tariff rates and/or small quotas, TRQs may be used to implement a trade regime as restrictive as the policies the TRQs were supposed to replace (see e.g. Herrmann *et al.*, 2001; Gervais and Rude, 2003). Switzerland can be seen as an eminent example for restrictive TRQs (Bureau *et al.*, 2019), as described in Section III.

Some previous studies have specifically analysed the effects of TRQs on domestic producer prices of agricultural products, such as Himics *et al.* (2020) for Swiss beef imports, Soon and Thompson (2019) for South Korean rice imports and Schmitz (2018) for US sugar imports. However, as we discuss in the next sub-section, few have examined the effects of TRQs, especially seasonal TRQs, on domestic producer prices intra-annual and disaggregated within the vegetable sector. Most computable equilibrium models, but also indicators of price support such as the OECD's price support estimate, fail to exhibit such a high level of detail.

To analyse the potential short-term effects of seasonal TRQs, intra-annual data are required. The topic of price volatility for agricultural products (mostly for grains and commodity futures) has been addressed frequently in recent decades (Fafchamps, 1992; Yang *et al.*, 2001; Lence and Hayes, 2002; Balcombe, 2009; Huchet-Bourdon, 2011; Wright, 2011; Chen and Villoria, 2019). Nevertheless, to our knowledge, only Abbott and Paarlberg (1998) explicitly discussed the effects of TRQs on price stability, arguing theoretically and empirically that frequent regime shifts from in-quota to out-of-quota tariffs may increase price volatility.

---

[3] For more literature on the theoretical effects of TRQs, see Skully (2001, 1999), Boughner *et al.* (2000) and Hranaiova and de Gorter (2005) for agricultural markets.

*Market price support for seasonal goods*

To the best of our knowledge, most evidence on seasonal tariffs involves the EU and Switzerland. The EU, which surrounds Switzerland, operates an entry price system (EPS) for vegetables and fruit. Although it is a seasonal system, it differs from the Swiss system. For many vegetables and fruits, the EU defines a product-tailored period around the harvest season in which more protective ad valorem tariff rates are set. The lower the price range within the price per kilo of imports falls, the higher the specific tariff rate that is applied in addition to the ad valorem tariff rate.[4] Four aspects are worth mentioning in view of our empirical study. First, there is no quota to restrict imports. Second, the phase with the higher tariffs for most vegetables starts several weeks before the harvest in Switzerland and ends later (see Appendix 1). Third, in terms of market size and geographical spread, the EU is a large customs union compared with Switzerland. Intra-EU trade is therefore relatively large compared with imports from outside, which are subject to the EPS. Regardless of the imports into the EU and the customs system applied, the producer prices in Switzerland's neighbouring countries are influenced by other EU countries with different climatic conditions. Fourth, even without the import protection duties, producer prices in the EU are significantly lower than those in Switzerland.

Goetz and Grethe (2009) analysed the importance of the EPS for imports of different fruits and vegetables in the EU. The two indictors of importance which they employ are based on the standard import values and the entry price; hence, they do not directly quantify the effects of the EPS on domestic producer prices. They conclude that the effects of the EPS are highest for artichokes, courgettes, cucumbers, lemons, plums and tomatoes (Goetz and Grethe 2009). Cioffi *et al.* (2011) analysed lemon and tomato prices for the EU from 2000 to 2007. They concluded that the 'resulting stabilization effect, as well as the support effect on EU domestic prices is rather small' (Cioffi *et al.,* 2011, p. 416). Martinez-Gomez *et al.* (2009), in their study of tomatoes, estimated that the abolishment of the EPS would reduce EU prices by up to 4.2%, which aligns with the findings of Antón-López and Muñiz (2007).

For Switzerland, Loi *et al.* (2016) analysed the impact of seasonal TRQs on the prices, imports and domestic production of potatoes, strawberries and apples. They found that the period protected by seasonal TRQs clearly restricted imports and increased consumer prices (Loi *et al.,* 2016). Regarding domestic producer prices, they pointed to an upward shift.

---

[4] The system has remained unchanged since 2013.

However, they did not quantify the effect. Finally, Hillen (2019) used weekly trade flows and trade costs data to estimate an extended parity bounds mode for Swiss and Italian tomatoes. She found that the probability of market inefficiency increased and market integration decreased with the seasonal TRQs (Hillen, 2019).

### III. INSTITUTIONAL BACKGROUND

*Agricultural market environment*

While Switzerland is, in general, a small, open economy, the agricultural sector, which contributes 0.7% to Switzerland's GDP and accounts for 3.4% of the country's employment (FSO 2019), stands apart from the rest of the economy. Swiss agriculture is traditionally characterised by a high level of market protection, with producer support estimates of 55% between 2016 and 2018 (OECD, 2019b). With a total factor productivity growth of less than 1.5% for the period from 2001 to 2016, Switzerland lags behind the average growth rate in the OECD (2019a). This lag can be explained partly by topographic conditions. Large areas of Switzerland are unsuitable for arable farming due to the hilly landscape (FSO, 2019). Prices paid to the farming sector are estimated to be 61% above global prices (OECD, 2016).[5] In particular, the vegetable market exhibits oligopolistic patterns (Chevalley, 2018). Few, often vertically integrated, processors and retailers exist (Logatcheva *et al.*, 2019). The interests of producers are aggregated and represented by producer organisations. The Association of Swiss Vegetable Producers (VSGP) is a key player and is also involved in the administration of import quotas, as described below.

*Seasonal TRQs*

Swiss agricultural policy is based on two cornerstones – a comprehensive system of direct payments and agricultural import regulation (Mann and Lanz, 2013; El Benni *et al.*, 2016). Virtually all imports of agricultural products into Switzerland are subject to tariffs or TRQs. While Swiss import tariffs amount to 2.3% on average for non-agricultural goods, they amount to 30.8% on average for agricultural goods (WTO, 2016).

Seasonal TRQs are implemented for fruits and vegetables. Out of season, there are no quantitative constraints on imports, and specific tariff rates in the range of $0–5 per 100 kilograms of vegetables, such as cherry tomatoes (FOAG, 2011), are applied.[6] For the majority

---

[5] A more extensive description of the Swiss agricultural market and the implications of agricultural trade regulations can be found in Gray *et al.* (2017).
[6] Preferential tariff rates for some country groups exist (see Jörin and Lengwiler, 2004; Khorana, 2008; Jörin, 2014).

of fruits and vegetables, a so-called administered – that is, protected – period exists.[7] These periods are specifically tailored to cover almost the entire domestic harvest season of each sort (see SWISSCOFEL, 2018 for an overview). During the protected phase, either lower in-quota tariff rates coinciding with the out-of-season rates or higher out-of-quota tariff rates are imposed. For instance, for cherry tomatoes, the out-of-quota rate amounts to at least 600 USD per 100 kilograms (FOAG, 2011).[8] As out-of-quota tariff rates are prohibitively high for most fruits and vegetables, the border can be considered de facto closed. This is also substantiated by econometric evidence by Hillen (2019) and Loi et al. (2016). Swiss authorities, more precisely the Federal Office of Agriculture (FOAG), either impose the lower in-quota tariff rates by defining a quantitatively restricted quota valid for a limited period or define a window of several days for which the lower in-quota tariff is applied without any quantitative restriction. FOAG decides whether to introduce import quotas of both types. These decisions can be revised up to twice a week during an administered period, and quotas can be set for times ranging from a couple days to the remaining duration of the administered period. A decision can be announced with a lead of half a week on imports.

Another distinct feature of this system is that the authorities decide on the size and time periods of the quotas in consultation with domestic producer associations and representatives of the processing and retail industry. While producers tend to favour smaller import volumes, processors and retailers usually advocate for higher volumes. The consumers, who are certainly the most price-conscious actors, are not directly represented in the negotiations. Overall, little is known about this repeated strategic game among the small number of players all commanding some market power.[9]

---

[7] For historical reasons, there exist an administered period and an effectively administered period. In the following, we always refer to the effectively administered period and use the term 'protected period' synonymously.

[8] More precisely, there are usually two out-of-quota tariffs. For each vegetable, authorities decide either to open no quotas at all or open them only for a given period of time. If they do not open a quota, the imposed tariff rate is higher than in the unprotected phase but slightly below the one applied if a quota was opened but filled. Both out-of-quota tariffs have the same economic effects because they usually suppress imports completely.

[9] The authorities usually expect producers, processors and retailers to agree on a joint proposal. The authorities aim at domestic market clearing and producer price stabilisation. Little is known publicly about how negotiations on TRQs may be influenced by other bilateral agreements between producers and processors, such as purchase agreements. Moreover, processors may have to buy certain shares of the annual domestic production to receive their shares in import quotas. A more detailed description can be found in Loi et al. (2016), especially in their supplementary material.

FOAG has the leeway to set the quotas within the protected period. However, the start and end dates of the protected phases are fixed in a federal ordinance (FOAG, 2016).[10] Short prolongations of the protected period would require broad political support, and long prolongations would even violate WTO rules. As political support for changes in the interest of the producer or in favour of more liberalised agricultural trade is not strong enough, the system of seasonal TRQs has remained practically unchanged since the mid-1990s.

## IV. Data and empirical approach

### Data

We aim to understand the effects of seasonal TRQs on the level and short-term volatility of producer prices in Switzerland. The time period of the study is from 2014 to 2019, which derives from the availability of price data for Switzerland.[11] Weekly producer prices for a wide range of vegetables issued by the VSGP are at the core of our study. There are two potential caveats to our data. First, the producer prices issued by the VSGP are recommended prices; therefore, these prices mostly reflect the prices for wholesale buyers, but some producers can sell at different prices. The producer prices provided by the VSGP usually contain packaging costs. Although the extent of packaging included in the price differs among the products, the extent remains constant over time for each product. The prices realised with these wholesalers are representative since there is a high market concentration in Switzerland.[12] The prices recommended by the VSGP are vital in the price setting as the organisation is also involved in the TRQ administration. Therefore, the prices reflect factual prices, on average, and they react quickly to changing market conditions.[13] The second caveat is that for most vegetables, price information is not available for the whole year. Our analysis of the weekly production quantities suggests that VSGP prices are issued whenever production takes place.[14]

The employed difference-in-differences approach requires price data from Switzerland's neighbouring countries as comparison groups. Therefore, we collected price information for

---

[10] A more readily accessible compilation of tariff rates by regime type and tariff line can be found in the guidelines issued by SWISSCOFEL (see e.g. SWISSCOFEL, 2018).

[11] While longer time series exist in principle, product reclassifications took place, and the inclusion of packaging costs may have changed over time. All these changes occurred at different points in time, possibly even during the main harvest season, that is, at points in time that would strongly affect our results if these changes were not correctly accounted for. However, no documentation exists describing exactly which amendment was carried out at specific times before 2014.

[12] As we describe below, time-invariant effects on prices are irrelevant for our econometric approach.

[13] These statements are based on direct communication with the producer organisation. The organisation collects fine-grained information on the market conditions for each vegetable on a weekly base. To test whether discrepancies between recommended and realised producer prices exist, a comparison with alternative data sources is required. Such data have become available only recently.

[14] Our data would have a truncation issue if price data were missing for weeks in which production was sold.

France from Franceagrimer, for Germany from the German Federal Office for Agriculture and Food (BLE) and for Italy from ISMEA (Istituto di Servizi per il Mercato Agricolo Alimentare). All three sources cover the same time period as the source used for Switzerland. The prices from France, Italy and Germany are the prices at the shipping stage (the first marketing stage). The comparison groups should by definition capture the prices that emerge under meteorologically and climatically comparable conditions, but in the absence of the Swiss TRQ system. Therefore, we restrict price data from neighbouring countries to regional markets close to the Swiss border and to prices for locally grown vegetables. For France, we consider the regions of Alsace-Lorraine, Auvergne Rhône-Alpes, Centre-Est and Roussillon; for Germany, we consider Frankfurt and Munich; and for Italy, we focus on the region north of Bologna. We carry out the econometric analysis at the most disaggregated level of the vegetable classification issued by the Swiss Centre for Vegetable Cultivation and of the Special Cultures (SZG). Seventy vegetables are subject to seasonal TRQs. The SZG collects two price series for most of these vegetables, one for conventionally produced vegetables and one for organically produced vegetables.

Corresponding producer price data from neighbouring countries are available for 35 products. The product classifications for Germany and Italy roughly correspond to the SZG classification. Especially for France, multiple varieties for each Swiss vegetable exist. We follow a data-driven approach and separately analyse all pairwise combinations of Swiss products with comparable varieties in the first step. In the second step, we perform common pre-trend tests, and we use these results to eliminate product combinations that do not satisfy the no-pre-trend condition. Some vegetables have longer production windows in Switzerland's neighbouring countries. We limit the sample to the weeks in which production takes place in Switzerland as we need a comparable market situation for the control group. In total, our analysis includes 12,093 available weekly price observations for Switzerland and 27,542 observations for neighbouring countries.

*Empirical approach*

We want to uncover the causal relationship between the protected period and vegetable prices in Switzerland. For this purpose, we apply a difference-in-differences approach. Hence, we compare the price differences between the protected and unprotected periods in Switzerland with the price differences occurring over the same time span in neighbouring countries. International price comparisons can suffer because product characteristics, such as product quality, differ between countries. However, under the assumption that prices both in Switzerland and the neighbouring countries would, on average, follow the same time trend in the absence of protected periods, constant differences in level net out, and the neighbouring countries serve as a control group for identifying this common trend. We expect the common trend assumption to hold only for treatment and control observations with similar observed covariates, which is weaker than imposing the assumption unconditionally.

Formally, we let $T$ be an indicator taking the value of 1 if the corresponding price observation falls into the administrated, meaning treated, period in which seasonal TRQs can be used to impede imports and 0 otherwise. Since we work with weekly data, some weeks can belong partially to the protected and the unprotected phases, which contaminates the pure effect of $T$. We therefore drop these observations from the sample.[15]

We denote by $Y$ either the standardised producer price $\tilde{P}$ or a measure of price volatility $\hat{P}$. $\tilde{P}$ captures the (absence of) price variations *across* the two phases (i.e. the unprotected versus the protected phase relative to the comparison group). $\hat{P}$ reflects the short-term or intra-phase volatility. To render weekly producer prices comparable between weeks, products and countries and across seasons, they were standardised by the average weekly producer price $\bar{P}_{s,i,c}$, where $s$ denotes the season, $i$ the product and $c$ the country at hand, while $w$ denotes weeks [16]: $\tilde{P}_{w,i,c} = \frac{P_{w,i,c}}{\bar{P}_{s,i,c}} * 100$. A season is defined as starting and ending in the middle of two administered periods.[17] Price volatility is operationalised by the absolute value of the percentage change in the price from week $w-1$ to $w$ (see e.g. Huchet-Bourdon, 2011):

$$\hat{P}_{w,i,c} = \left| \frac{P_{w,i,c}}{P_{w-1,i,c}} - 1 \right|$$

The binary variable $D$ denotes the assignment to either the treatment group ($D=1$) for a given Swiss vegetable or the control group ($D=0$) for the corresponding vegetable from a neighbouring country. Applying the potential outcome notation (Rubin, 1974), we let $Y(1), Y(0)$ denote the potential outcomes (i.e. the prices hypothetically realised with and without TRQs). We are interested in the average treatment effect on the treated (ATET) in the protected period, defined as the difference in the potential outcomes for products exposed to TRQs in the protected period:

$$ATET_{D=1,T=1} = E[Y(1) - Y(0)|D = 1, T = 1]$$

---

[15] Hence, for level variables, we drop the start and end weeks for the administered period from the sample whenever the period does not start on a Monday or end on a Sunday, respectively. For variables that reflect changes from week $w-1$ to $w$, we always removed the one or two weekly observations around the transition from the unprotected to protected phase (and vice versa), which are based on weekly prices from both phases.

[16] This standardisation was chosen for better readability and because the weekly produced quantities required to compute weighted averages are unknown for some products.

[17] The start and the end of the administered periods were chosen by the authorities such that they cover the main harvest period. Therefore, for most products, no production takes place around the start and end dates of the season, according to our definition; hence, the choice of the season start is of little relevance in the analysis.

Besides the already-mentioned common trend assumption that needs to hold conditional on covariates, identification requires that the so-called stable unit treatment valuation (SUTVA) be satisfied, implying that the TRQs in Switzerland have no spill-over effects on prices in neighbouring countries and that there are no anticipatory effects, implying that the TRQ does not affect the price setting before it enters into force (i.e. in the non-protected period). Concerning the plausibility of SUTVA, we note that spill-overs of the Swiss trade regime on the price setting in the control group could lead to violations of this assumption. The magnitude of such general equilibrium effects depends on how important the Swiss market is for producers in neighbouring countries. We examined Swiss imports from neighbouring countries and concluded that these effects could be neglected. For products imported from France and Germany, the Swiss market is of rather low importance as the share of products exported to Switzerland from France, Germany and Italy is less than 20%, for most products, except asparagus (35%) (see Appendix 2). While the dependency of Switzerland on imports from Italy is somewhat higher for certain vegetables, Switzerland, as a small country, is still unlikely to have an important impact on Italian producer prices.

Furthermore, common support must hold in the sense that, for treated units in the protected period, comparable units with similar covariates exist in the following three populations: treated units in the unprotected period, controls in the protected period and controls in the unprotected period. Under these assumptions, the ATET is obtained by (i) taking the difference in average price differences (between protected and unprotected periods) across Swiss and neighbouring observations with similar covariates and (ii) averaging this difference over the covariate distribution among the treated in the protected period:

$$ATET_{D=1,T=1} = E[E[Y|D = 1, T = 1, X] - E[Y|D = 1, T = 0, X]$$
$$- \{E[Y|D = 0, T = 1, X] - E[Y|D = 0, T = 0, X]\}|D = 1, T = 1]$$

$X$ represents observed covariates on which we condition to make the common trend assumption more plausible. We include fixed effects for seasons in all subsequently presented specifications, while we pool all data from the years and all weekly observations by phases.[18] Concerning the plausibility of the no-anticipation assumption, we note that the TRQ system has been almost unchanged for more than two decades, as outlined in the institutional setting. The start and end dates of the administered periods cannot be prolonged, and the system is mandatory for producers and importers of vegetables that are subject to the system of seasonal TRQs. For this reason, the actors are aware of these regulations such that anticipatory price effects in the pre-treatment period may be an issue. Therefore, we test the common trend assumption by means of the abovementioned pre-trend tests.

---

[18] We performed robustness checks with controls for weeks and months as well. These results are available from the authors on request. For the details, see Appendix 4.

We estimate the effect of the administered period product-by-product using difference-in-differences based on inverse probability weighting (Abadie, 2005; Lechner, 2011). This approach reweights treated observations in the unprotected period and control observations in both the protected and unprotected periods such that the respective covariate distributions of the three groups match the distribution of the treated observations in the protected period for which the effect is estimated. Reweighting is based on the inverse of the propensity score, which corresponds to the conditional probability observed in a specific treatment group and period as a function of the observed covariates. To this end, we employ the *didweight* function in the *causalweight* R package (Bodory and Huber, 2019). We also impose common support in the sample by dropping observations with propensity scores of being treated in the protected period larger than 0.95 or 0.99 relative to other groups (in a different treatment state and/or time period). The reason for the imposition is that such large propensity scores imply that for specific treated observations in the protected period, observations with similar covariates in the groups of treated in the unprotected period, controls in the protected period and/or controls in the unprotected period are rare or non-existent. Imposing common support avoids issues by using non-comparable observations (in terms of covariates) across groups (e.g. a high variance of the estimator) at the cost of reducing the analysis to a subset of the data, which could reduce external validity.

In the results section, we focus on the ATET for the whole protected period. In addition, we also estimated the ATET for each week of the protected period separately. Furthermore, we conducted placebo tests for differential pre-treatment trends across treated and control observations in the unprotected periods. That is, we considered observations four and three weeks prior to the start of the protected phase per vegetable as the pre-treatment period and observations in the last two weeks prior to the start of the protected phase as the placebo-treatment period in order to test whether the trends diverged.[19] As our analysis aims at quantifying the effect of Swiss TRQs on vegetable prices relative to the EU, it is important to note that the EU imposes tariffs on some vegetable groups, too, as outlined in Appendix 1. While these EU tariffs are less stringent than the Swiss ones, they nevertheless imply that our estimates for these respective vegetables constitute a lower bound on the effect of a hypothetical comparison of Swiss TRQs vs. no TRQs at all.

---

[19] Since we use only two weeks per period, we cannot include seasonal fixed effects. Analogously, we estimated the treatment effects separately on a rolling biweekly basis during the protected phase. For this purpose, we always use the last two weeks before the start of the protected phase as the pre-treatment phase. As the number of observations is rather low for weekly estimates and the pre-trend tests, we do not present the results in the text. However, the results are available from the authors on request.

## V. RESULTS

### *Descriptive evidence*

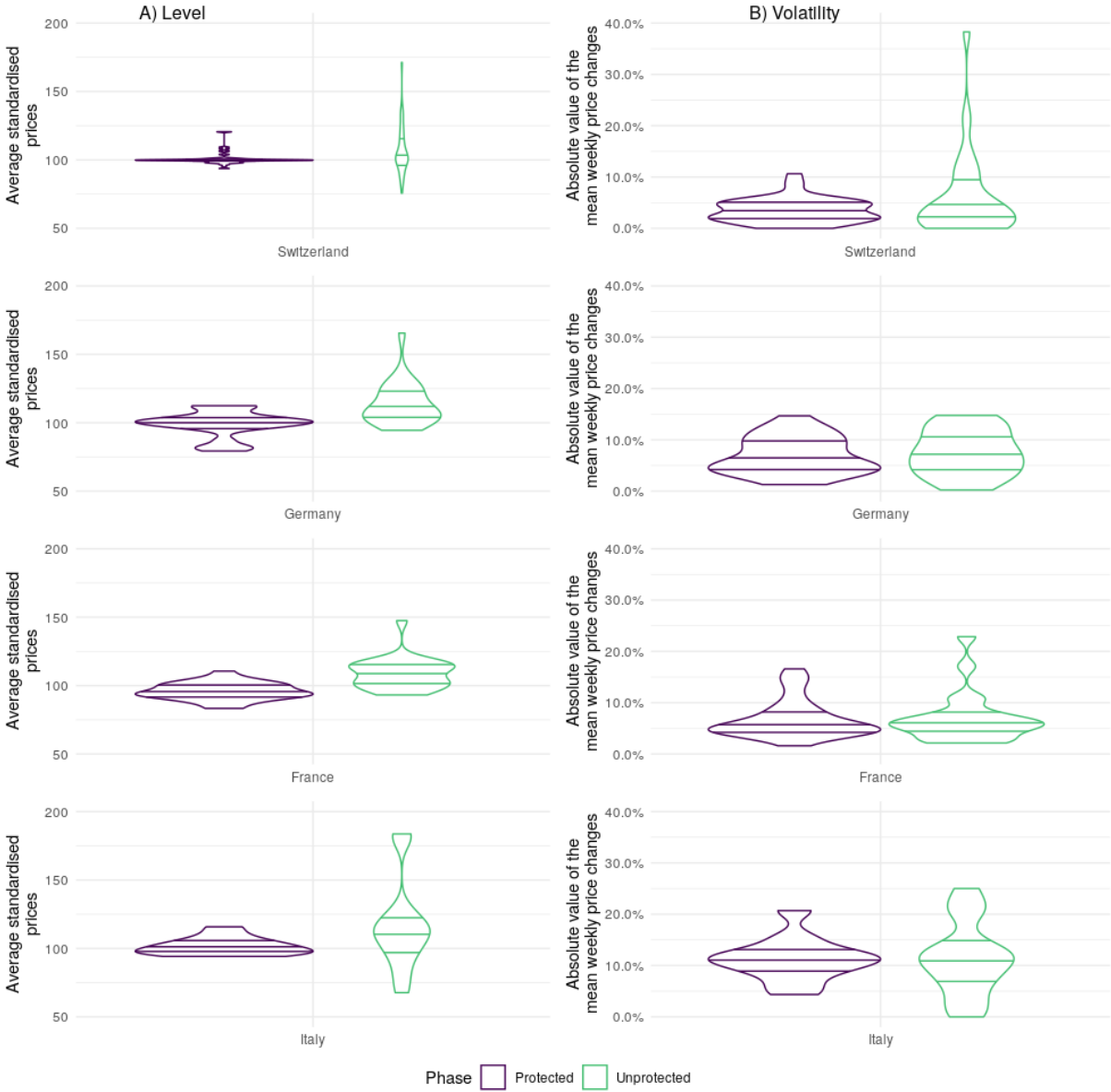

Figure 1 illustrates all data on which the subsequent econometric analysis is based by means of violin plots. Violin plots visualise the distribution of variables like box plots; they also show the kernel probability density of the data. The three lines in the plots mark the first, second (median) and third quartiles. Prices are depicted separately for the Swiss administered period, in which domestic production is protected by trade barriers, and the unprotected period. The units of observation in the plots are product- and period-wise averages for the two outcome variables – price level in panel A and volatility in panel B.

As price levels are standardised by product, the medians of the protected and the unprotected periods naturally floated around 100 for each product. For Germany and France, it is apparent that prices during the Swiss protected phase are, on average, lower (98 for Germany and 95 for France) than in the unprotected phase (115 for Germany and 109 for France) as supply expands quickly relative to demand during harvest. For Italy, the same is true (108 in the unprotected and 97 in the protected phase). However, the overall picture is dominated by some vegetables that exhibit extreme price reactions. Prices are more dispersed for the unprotected phase for all countries.[20] For Switzerland, the graph shows two distinct results. First, price levels are very stable. The average standardised price is at 101 for the protected phase and 105 for the unprotected phase. Second, prices in the protected phase are strongly clustered around 100. This result also reflects that, compared with neighbouring countries, Swiss production is more concentrated in the protected than the unprotected phase.

---

[20] This pattern can, in part, be explained by the standardisation by seasons and products in combination with the number of observations. As more weekly price observations are available for the protected phase, which covers the main harvest period, the average weekly price in the protected phase must be closer to the average weekly price of the whole season.

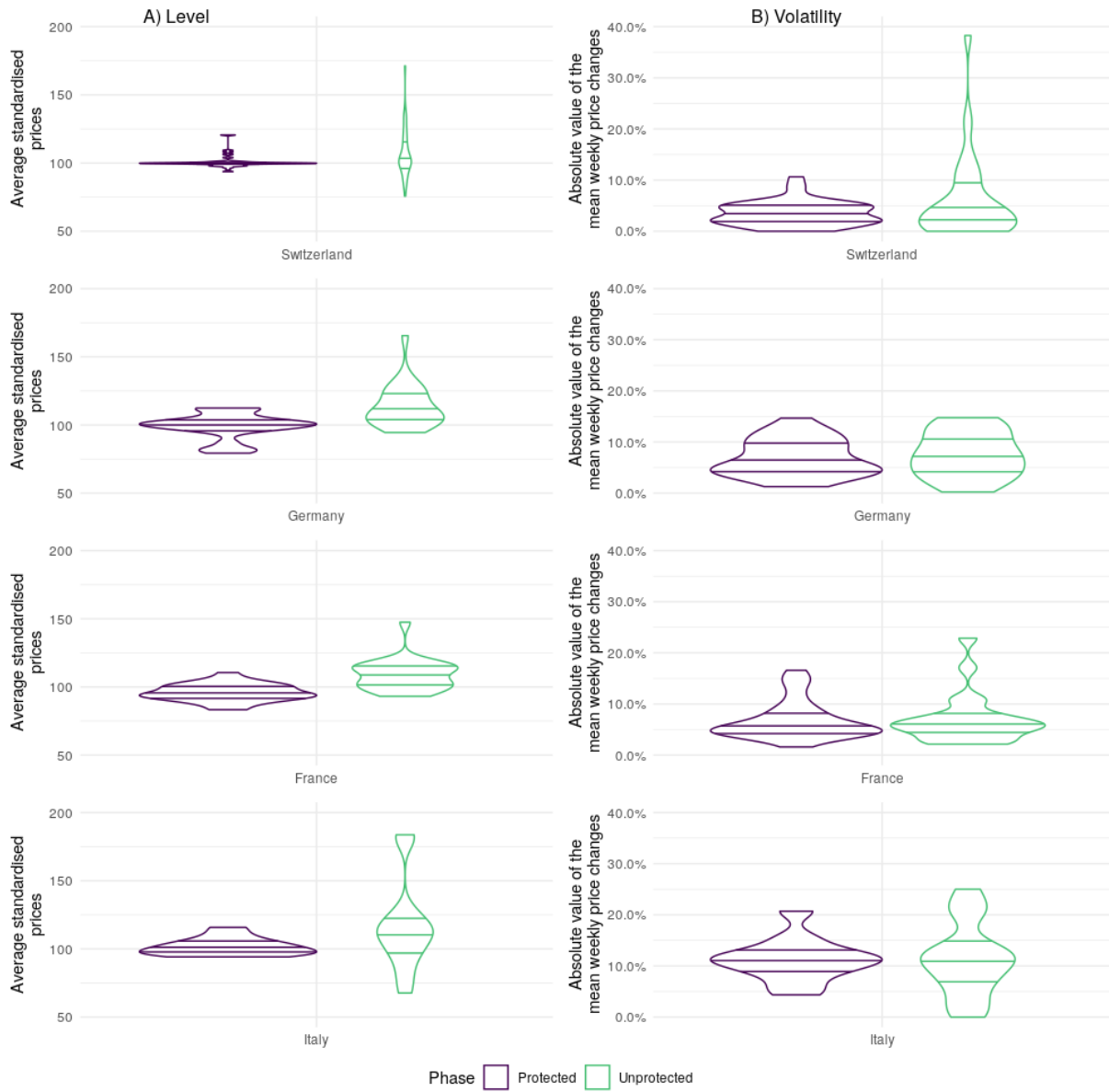

*Figure 1. Vegetable price levels and volatility in Switzerland and neighbouring countries*

Turning to the volatility measured as week-to-week price change, we find that price volatility is similar within the protected and unprotected phases in France (protected: 6.9%; unprotected: 6.3%), Germany (protected: 6.4%; unprotected: 7.6%) and Italy (protected: 11.4%; unprotected: 10.8%). The graph shows that Switzerland generally has lower price volatility. The volatility is also lower in the protected phase (2.8%) than in the unprotected phase (3.4%).

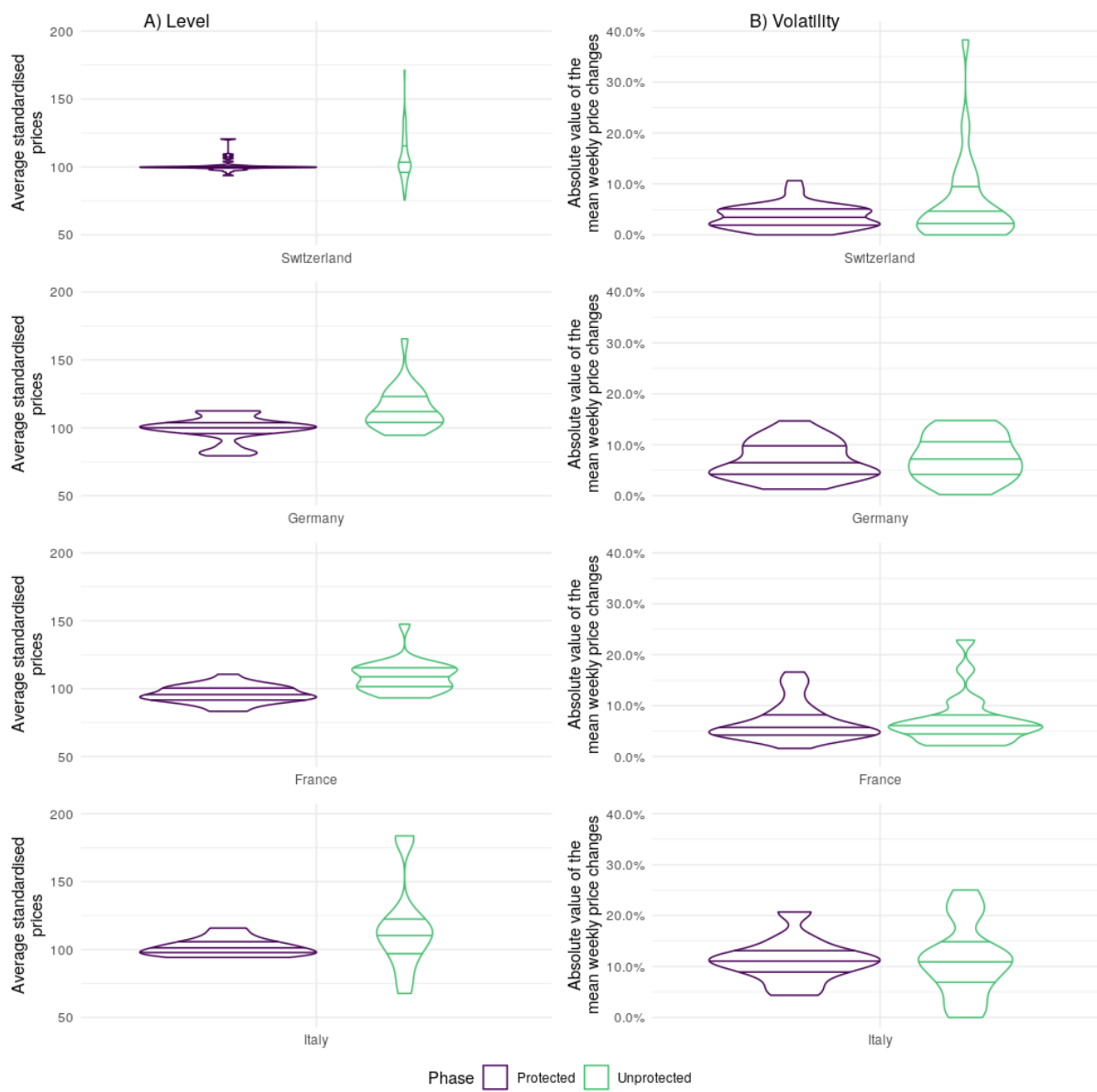

Figure 1 shows that Switzerland exhibits a price level and volatility pattern similar to those of its neighbours. However, it is evident that in neighbouring countries, prices change in the course of a season, thus highlighting the need for a difference-in-differences approach at the level of individual vegetables.

*Price effects of seasonal TRQs*

Table 1 contains the difference-in-differences estimates for the effects of seasonal TRQs for vegetables in Switzerland on producer price levels and volatilities. Observations with propensity scores of being treated in the protected period that are larger than 0.95 are discarded to ensure common support. As can be seen by the blanks in the table, some effects could not be estimated. For 18 out of 35 vegetables, only a few harvest days exist outside the protected phase such that there are no observations to compare in the unprotected phase.[21] As the production periods for conventional and organic vegetables differ, the availability of results also differs between the two production methods. We list the products for which the number of harvest weeks outside the protected period is too small for estimations in Appendix 3.

Regarding prices for conventional vegetables, we find that seasonal TRQs lead to higher prices for most vegetables, while the effects on weekly price volatilities are mixed and rather modest. The effect of the protected phase on the producer price is significantly positive except for Batavian lettuce, leeks, red oakleaf lettuce and courgettes. The significant price increases range from 7.5 for Brussels sprouts to 41 for truss tomatoes and 91.2 for round tomatoes.[22] Owing to the price standardisation, these estimates correspond to percentage changes relative to the average weekly price in Switzerland.

In the protected phase, the price volatilities of conventional vegetables are approximately five percentage points higher for Brussels sprouts and leeks and two to three percentage points for cherry tomatoes compared with the unprotected phase. By contrast, price volatilities are 20 percentage points lower for (regular) tomatoes and approximately five percentage points lower for truss tomatoes in the protected phase.

The effects of seasonal TRQs are less pronounced and more heterogeneous for organically produced vegetables. The prices for cauliflower, spinach, eggplant and most kinds of tomatoes are higher in the protected phase than in the unprotected phase. If prices are higher, the increase tends to be even more pronounced than for conventional production, as the price increase of 90.8 percentage points for round tomatoes shows. However, prices for fennel, greenhouse cucumbers and courgettes decreased by 17–34 percentage points. For Batavian lettuce, Brussels sprouts, cherry tomatoes, eggplant, leeks, spinach, tomatoes and courgettes, we observed an increased price volatility of up to seven percentage points in the protected phase, while price volatility is lower for round tomatoes only compared with Italy when comparing the protected phase with the unprotected phase. Appendix 4 presents DID results based on a linear specification using OLS (Table A3) as well as based on inverse probability weighting with propensity scores of being treated in the protected period that are lower than 0.99 (Table A4). The estimates by and large confirm the findings in Table 1.

---

[21] The calculation of $t$ weekly price changes requires data from $t + 1$ weeks. One or two weekly price changes per regime shift from protected to unprotected, and vice versa, need to be removed from the sample as these observations would mix weekly prices from both regimes. For this reason, the number of observations is too low to estimate the effects of the TRQs on volatility for some vegetables, such as eggplant or white radishes, while price level estimates are shown.

[22] Effects close to 50% may look large. However, the weekly prices are not weighted by quantities; hence, low prices may apply to a very small production quantity in the unprotected phase, and production time periods may differ. Producer prices are usually disseminated only when production takes place.

Our results show that many vegetable price levels and volatilities behave differently in the protected phase compared with the unprotected phase and compared with neighbouring countries. However, whether the causal effect of the seasonal TRQ system is properly identified depends on whether the assumptions described in Section IV hold. In particular, the method requires a parallel trend assumption. To check the validity of this assumption, as mentioned in Section IV, we used placebo tests for differential time trends in the prices of treated and control groups in the unprotected period; see Table 1. For most products, we do not find indications that Swiss producer prices and those from neighbouring countries diverge even before the administered period starts. However, as we only pool data within two weeks for this test, the number of observations is rather low, and in the case that a missing observation occurs in one or two years for either the treatment or the control group, the test cannot be performed.

Furthermore, and as acknowledged by Roth (2019), the power of such placebo tests for finding violations of the common trend assumption might be rather low (but see Jaeger et al. [2020] for an example of where such tests do overturn the initial empirical findings). However, a violation in a relatively small number of tests does not imply a violation of the common trend assumption. This is due to the multiple hypothesis testing issue that the probability of spuriously rejecting a correct null hypothesis of common trends in some of the tests increases with the number of tests conducted. For these reasons, Roth (2019) encourages researchers to scrutinise the plausibility of the common trend assumption in the given economic context. We argue that the fact that the treatment and control groups in our data reflect geographically and culturally closely related regions is in favour of the common trend assumption, in particular after standardising the prices and de-trending volatilities.

Six out of the 17 vegetables in Table 1 are subject to seasonal market protection in the EU (see additional details in the notes for Table 1 and also in Appendix 1). Empirical investigations suggest that these measures *increase* domestic prices in the EU by up to 4.2 % for tomatoes and 8.3% for lemons (Martinez-Gomez *et al.*, 2009), while the stabilisation effect appears rather small (Cioffi *et al.,* 2011). We found that the magnitudes of the Swiss TRQ effects on producer prices are much higher than the suggested effects in the EU. As the policies in the EU and Switzerland both seem to increase domestic prices, the effect of the Swiss regime might be underestimated if the EU tariffs enter into force in the time span within which we chose to measure the effect of Swiss TRQs. Our estimates are then a lower bound on the true effect of the Swiss TRQs. The shares of statistically significant positive findings are 92% and 66% for conventional and organic price levels, respectively, and 57% and 77% for conventional and organic price volatilities, respectively. The positive price shifts and the increase in price volatility strongly outnumber the opposite effects.

*Table 1. Price effects of seasonal TRQs on vegetable prices*

| | | Conventional | | | | Organic | | | |
|---|---|---|---|---|---|---|---|---|---|
| | | Level | | Volatility | | Level | | Volatility | |
| Vegetable | Comparison | Effect (standard error) | Pre-trend | Effect (standard error) | Pre-trend | Effect (standard error) | Pre-trend | Effect (standard error) | Pre-trend |
| Batavian lettuce | France | 19.7 (13.9) | | 0.007 (0.034) | | -6.29 (12) | | 0.046 (0.014)*** | |
| Broadleaf endive | France | | | | | 20.1 (14.7) | | | |
| Broadleaf endive | Germany | | | | | 3.01 (3.82) | | | |
| Brussels sprout | Germany | 7.54 (3.86)* | | 0.049 (0.022)** | | -7.29 (5.54) | | 0.059 (0.012)*** | |
| Cauliflower | Germany | | | | | 60.7 (14.3)*** | | | |
| Cauliflower | Italy | | | | | 1.87 (22.8) | | | |
| Cherry tomato | Germany | 35.7 (6.69)*** | 16.6 (4.48)*** | 0.028 (0.017)* | -0.044 (0.044) | 10.6 (4.57)** | 8.84 (4.25)** | 0.064 (0.013)*** | |
| Eggplant | Germany | 25.6 (5.05)*** | | 0.021 (0.028) | | 28.1 (16.5)* | | 0.0642 (0.0311)** | |
| Eggplant | Italy | 29.6 (7.45)*** | | | | 29.3 (13.8)** | | | |
| Florence fennel | Italy | | | | | -22.2 (8.57)*** | | | |
| Greenhouse cucumber | France | 20.6 (3.05)*** | 8.86 (6.44) | 0.024 (0.018) | -0.029 (0.029) | -18.9 (5.17)*** | -2.76 (6.97) | -0.006 (0.020) | 0.061 (0.038) |
| Greenhouse cucumber | Germany | 19.8 (3.88)*** | 9.75 (6.42) | 0.025 (0.026) | 0.0131 (0.035) | -20.5 (6.44)*** | -0.239 (7.61) | -0.031 (0.028)* | 0.12 (0.042)*** |
| Leek | France | 2.53 (11.4) | -8.84 (10.4) | 0.026 (0.013)** | -0.007 (0.017)* | 4.24 (11) | -10.6 (11.3) | 0.014 (0.0127) | 0.010 (0.016) |
| Leek | Germany | 1.46 (10.8) | -0.652 (6.84) | 0.007 (0.011) | 0.002 (0.016) | 2.35 (9.05) | -0.454 (7.05) | 0.001 (0.011) | 0.028 (0.0125)** |
| Leek | Italy | -31.2 (17.9)* | 1.6 (10.3) | 0.054 (0.017)*** | -0.043 (0.056) | -28.8 (22.3) | | 0.052 (0.010)*** | |
| Oakleaf lettuce (green) | France | 25.6 (15.5)* | | | | | | | |
| Oakleaf lettuce (red) | France | 18.9 (12.8) | | -0.001 (0.01) | | | | | |
| Radish (white) | France | 35.7 (10.7)*** | | | | | | | |
| Slicing cucumber | Italy | -29.2 (19.2) | | | | -13.4 (15.4) | | | |
| Spinach | France | | | | | 14.1 (19.3) | | 0.029 (0.011)*** | |
| Spinach | Italy | | | | | 51.4 (10.9)*** | | | |
| Tomato | Germany | 36.2 (5.72)*** | | 0.046 (0.035) | | 17.7 (4.73)*** | | 0.058 (0.033)* | |
| Tomato | Italy | 91.2 (1.6)*** | | -0.205 (0.01)*** | | 90.8 (7.37)*** | | -0.095 (0.0053)*** | |
| Truss tomato | France | 33.3 (5.99)*** | 28.7 (8.2)*** | -0.050 (0.027)* | -0.062 (0.047) | -2.4 (3.38) | | -0.020 (0.037) | |
| Truss tomato | Germany | 41 (1.25)*** | 11.9 (5.03)** | 0.010 (0.013) | 0.025 (0.029) | 14.7 (4.69)*** | | 0.019 (0.014) | |
| Courgettes/summer squash | Germany | 2.1 (11.4) | | 0.015 (0.0154) | | -34 (20.9) | | 0.098 (0.060)* | |
| Courgettes/summer squash | Italy | -5.07 (23.6) | | 0.149 (0.121) | | -33.8 (17.5)* | | 0.015 (0.020) | |

**Notes:** Difference-in-differences estimates were obtained using inverse probability weighting. The column 'effects' denotes the ATET for the protected phase either regarding the standardised price (a price of 100 corresponds to the average weekly price per season of the respective vegetable) or the relative weekly price change. For the main effect estimation, data for each vegetable were pooled by phases (protected/unprotected), and seasonal and biweekly fixed effects were employed. For the pre-trend test, observations from four and three weeks before the protection started were employed as the pre-period, while observations from the last two weeks before the protection started were used as a pseudo treatment period. No covariates were used in the pre-trend test estimation. ***, **, * denote $p$-values smaller than .01, .05 and .1, respectively. In the EU, greenhouse cucumbers and all types of tomatoes are protected by seasonal entry prices and seasonal tariffs, whereas cauliflower has only seasonal ad valorem tariffs, and courgettes have only seasonal entry prices (see Appendix 1). Observations with propensity scores of being treated in the protected period that are larger than 0.95 are dropped from the sample to ensure common support. Appendix 4 presents the results of OLS (Table A3) and DID (Table A4) with propensity scores lower than 0.99.

## VI. DISCUSSION AND CONCLUSION

Few studies have investigated the effects of seasonal TRQs on domestic producer prices on vegetable markets, and those have looked only at certain selected products. Vegetable markets are characterised by large heterogeneity in the production process among vegetables and by the importance of short-term and regional influences on production. We contribute to the literature by using a difference-in-differences approach based on weekly producer data from Switzerland and neighbouring countries to estimate the effects of the protected phases on domestic producer prices for vegetables in Switzerland.

Our application focuses on seasonal tariff rate quotas and is rooted in some Swiss specificities. However, it is worth pointing out the generalisability of the method. While other countries, such as Norway, the US, the EU and Japan, also apply seasonal TRQs (Hillen, 2019; Hallam *et al.,* 2004; Johnson, 2017), the approach is well suited to quantify the effects of other temporally and spatially limited agricultural policy measures. These range from fertiliser or fuel subsidies in developing countries (Holden, 2019; Adetutu and Weyman-Jones, 2019) or labour programmes in the EU and Canada that include seasonal taxes for agricultural workers (OECD, 2020) to local seasonal water quality programmes (Brainerd and Menon, 2014). One of the few prerequisites is the existence of control regions, that is, countries or even sub-federal entities that are not subject to the policy of interest and which have otherwise comparable conditions and the necessary statistical data. Depending on the variables and the treatment of interest, these conditions may not only be met by highly developed and landlocked countries such as Switzerland.

From a methodological perspective, three properties of our dataset need to be discussed. First, the availability and representativity of observations outside the protected phases are key for our econometric approach. The protected phases are designed to cover the main harvest period of the vegetables. Hence, the low number of production weeks outside the protected phase does not come as a surprise. While the low availability of data in the unprotected phase reduces the number of vegetables for which we can perform an econometric analysis from 35 to 17, the analysis generated informative insights for our assumption that the production conditions evolve similarly from the unprotected to the protected phase in Switzerland and its neighbouring countries. The performed pre-trend tests mostly support this assumption. Second, the time period of five years limits the use of covariates and the estimation of weekly effects. A longer time horizon could compensate for the relatively low number of observations per season in the unprotected phase. Third, spatial conditions have a strong impact on the production of vegetables, and these conditions can vary locally. While data for vegetable production with a higher spatial resolution and closer to the Swiss border are unavailable, such data could increase the risk that the SUTVA is violated. This would be the case if the measures of market protection in Switzerland had spill-overs on the imports from neighbouring regions. However, this study focuses on price effects and not on production. The prices from neighbouring regions depend on demand on national or even larger markets rather than local markets. As Switzerland is a small country, imports to Switzerland or the absence of these imports have only a minor impact on international markets.

Our analysis clearly showed that seasonal TRQs increase prices for vegetables by pushing the standardised weekly prices, often by 25% and in some cases even by more than 90% above the prices in neighbouring countries, as the evidence for prices on Swiss and Italian tomatoes shows. Compared with the average market price support of 55% for the agricultural sector in Switzerland, according to OECD (2019b) estimates, we observe that Swiss producers of some vegetables, such as tomatoes and cauliflower, receive much stronger *temporal* support than the producers in neighbouring countries, while Swiss producers of other vegetables, such as leeks, may get lower comparative price support. However, how much this temporary support contributes to the overall support of the vegetable markets is beyond the scope of this paper.

It is not surprising that the magnitudes of the TRQ effects on producer prices we find in Switzerland are much higher than the suggested effects in the EU (Martinez-Gomez *et al.*, 2009) as the Swiss system allows the setting of import duties at a magnitude that has a prohibitive effect on imports from most countries, while the EU imposes tariffs only for imports from non-EU countries. At the same time, our analysis did not show systematic effects of the seasonal TRQs on the week-to-week price changes when comparing the unprotected phase with the protected phase. Our results resemble the effects of EPS found in the EU (Cioffi *et al.*, 2011, pp. 416) – the increase of the EU domestic average prices due to EPS and the decrease of price standard deviations. Since the Swiss seasonal TRQs may also generate the incentive to align Swiss production to a narrow time window, a higher short-term price variance would have been conceivable as a side effect. As the price volatility is generally lower in Switzerland than in neighbouring countries, it can be hypothesised that the TRQ system contributes to this overall price stability.

Finally, since we apply a common methodology to study a large array of vegetables in a common setting, we are able to detect considerable heterogeneity in the effects of seasonal TRQs between vegetables. While we were only partially able to explain the determinants, we found larger price increases for conventional production due to TRQs. Fostering organic production is a central goal of not only Switzerland's agricultural policy (Hirschi and Huber, 2012). Denmark and Sweden – which joined Switzerland in the top three organic food consumers per capita in 2018 (FiBL, 2020) – but also other EU countries; Asian countries, such as Bhutan (FAO, 2012); and African countries, such as Uganda (see the details in IFOAM [2018]) are increasingly orienting their agricultural policy towards organic production. Hence, understanding the differential effect of policies on organic production is an important avenue for further research.

## VII. REFERENCES

Abadie, Alberto. 'Semiparametric difference-in-differences estimators', *Review of Economic Studies*, Vol. **72**, (2005) pp. 1–19.

Abbott, Philip C. 'Stabilisation policies in developing countries after the 2007-08 food crisis', in *Agricultural Policies for Poverty Reduction*, Jonathan Brooks, ed. (Wallingford: CABI, 2012), 109–168.

Abbott, Philip C. and Paarlberg, Philip L. 'Tariff rate quotas: structural and stability impacts in growing markets', *Agricultural Economics*, Vol. **19**(3), (1998) pp. 257–267.

Adetutu, O. Morakinyo and Weyman-Jones, Thomas G. 'Fuel subsidies versus market power: is there a countervailing second-best optimum?' *Environmental & Resource Economics,* Vol. **74**, (2019) pp. 1619–1646.

Aksoy, M. Ataman and Beghin, John C. *Global Agricultural Trade and Developing Countries.* (Washington D.C.: World Bank, 2005).

Antón-López, Jesus and Muñiz, Ignacio A. 'Measuring domestic implications of tariff cuts under EU entry price regime', in *103rd Seminar of the European Association of Agricultural Economists* (2007).

Balcombe, Kelvin. 'The Nature and Determinants of Volatility in Agricultural Prices'. MPRA Paper No. 24819 (2009).

Bodory, Hugo and Huber, Martin. 'The "causalweight" package'. 2019. https://cran.r-project.org/web/packages/causalweight/vignettes/bodory-huber.pdf.

Bonferroni, Carlo E. *Teoria statistica delle classi e calcolo delle probabilità.* (Firenze: Libreria Internazionale Seeber, 1936).

Boughner, Devry S., de Gorter, Harry, and Sheldon, Ian M. 'The economics of two-tier tariff-rate import quotas in agriculture', *Agricultural and Resource Economics Review*, Vol. **29**(1), (2000) pp. 58–69.

Brainerd, Elizabeth and Menon, Nidhiya. 'Seasonal effects of water quality: the hidden costs of the green revolution to infant and child health in India', *Journal of Development Economics,* Vol **107** (C), (2014) pp. 49–64.

Bureau, Jean-Christophe, Guimbard, Houssein, and Jean, Sebastien. 'Agricultural trade liberalisation in the 21st century: has it done the business?' *Journal of Agricultural Economics*, Vol. **70**(1), (2019) pp. 3–25.

Chen, Bowen and Villoria, Nelson B. 'Climate shocks, food price stability and international trade: evidence from 76 maize markets in 27 net-importing countries', *Environmental Research Letters*, Vol. **14**(1), (2019), 014007.

Chevalley, Majorie. 'The fruit and vegetable market in Switzerland: overview of the market and access information for international trading companies'. 2018. http://www.swisscofel.ch/wAssets/docs/news/Fruit-and-vegetable-market-in-Switzerland_angepasst-Maerz-2018.pdf.

Cioffi, Antonio, Santeramo, Fabio G., and Vitale, Cosimo D. 'The price stabilization effects of the EU Entry price scheme for fruit and vegetables', *Agricultural Economics*, Vol. **42**(3), (2011) pp. 405–418.

El Benni, Nadja, Finger, Robert, and Meuwissen, Miranda P. M. 'Potential effects of the Income Stabilisation Tool (IST) in Swiss agriculture', *European Review of Agricultural Economics*, Vol. **43**(3), (2016) pp. 475–502.

Fafchamps, Marce. 'Cash crop production, food price volatility, and rural market integration in the third world', *American Journal of Agricultural Economics*, Vol. **74**(1), (1992) pp. 90–99.

FAO (Food and Agriculture Organization of the United Nations). *Bhutan. Agricultural Sector Review. Volume 1, Issues, Institutions and Policies.* (Rome: FAO, 2012).

FiBL (Research Institute of Organic Agriculture). *European organic market grew to 40.7 billion euros in 2018*. Media release February 12, 2020.

FOAG. *Verordnung über die Einfuhr von landwirtschaftlichen Erzeugnissen: AEV 916.01* (2011).

FOAG. *Verordnung des BLW über die Festlegung von Perioden und Fristen sowie die Freigabe von Zollkontingentsteilmengen für die Einfuhr von frischem Gemüse und frischem Obst: VEAGOG 916.121.100* (2016).

FSO (Swiss Federal Statistical Office). *Land- und Forstwirtschaft: Panorama* (Neuchâtel: Federal Statistical Office, 2019).

GATT. 'Agreement on agriculture', 1994. https://www.wto.org/english/docs_e/legal_e/14-ag_01_e.htm.

Gervais, Jean-Philippe and Rude, James I. 'Some unintended consequences of TRQ liberalization', *Journal of Agricultural & Food Industrial Organization*, Vol. **1**(1), (2003).

Goetz, Linde and Grethe, Harald. 'The EU entry price system for fresh fruits and vegetables – paper tiger or powerful market barrier?' *Food Policy*, Vol. **34**(1), (2009) pp. 81–93.

Gray, Emily, Adenäuer, Lucie, Flaig, Dorothee, and van Tongeren, Frank. *Evaluation of the Relevance of Border Protection for Agriculture in Switzerland.* (Paris: OECD Publishing, 2017).

Hallam, David, Liu, Pascal, Lavers, Gill, Pilkauskas, Paul, Rapsomanikis, George, and Claro, Julie. 'The market for non-traditional agricultural exports,' FAO Commodities and Trade Technical Paper, No. 3, 2004, Chapter 3.2.

Herrmann, Roland, Kramb, Marc, and Mönnich, Christina. 'Tariff rate quotas and the economic impact of agricultural trade liberalization in the World Trade Organization', *International Advances in Economic Research*, Vol. **7**, (2001) pp. 1–19.

Hillen, Judith. 'Market integration and market efficiency under seasonal tariff rate quotas', *Journal of Agricultural Economics*, Vol. **70**(3), (2019) pp. 859–873.

Himics, Mihaly, Listorti, Giulia, and Tonini, Axel. 'Simulated economic impacts in applied trade modelling: a comparison of tariff aggregation approaches', *Economic Modelling*, Vol. **87,** (2020) pp. 344–357.

Hirschi, Christian and Huber, Robert. 'Ökologisierung der Landwirtschaft im agrarpolitischen Prozess', *Agrarforschung Schweiz*, Vol. **3**(7-8), (2012) pp. 360–365.

Holden, Stein T. 'The economics of fertilizer subsidies', Centre for Land Tenure Studies Working Paper, No. 9/18, 2019.

Hranaiova, Jana and de Gorter, Harry. 'Rent seeking with politically contestable rights to tariff-rate import quotas', *Review of International Economics*, Vol. **13**(4), (2005) pp. 805–821.

Huchet-Bourdon, Marilyne. 'Agricultural commodity price volatility: an overview', OECD Food, Agriculture and Fisheries Working Papers, No. 52, 2011.

IFOAM. 'Change for good. Organics International Annual report', (2018).

Jaeger, David A., Joyce, Theodore J., and Kaestner, Robert. 'A cautionary tale of evaluating identifying assumptions: did reality TV really cause a decline in teenage childbearing?' *Journal of Business & Economic Statistics,* Vol. **38**(2), (2020) pp. 317–326.

Johnson, Renée. 'Efforts to address seasonal agricultural import competition in the NAFTA renegotiation', Congressional Research Service, 7-5700, 2017.

Jörin, Robert. 'Improving market access: the role of auctions in converting tariff-rate quotas into single tariffs' *Review of Agricultural and Applied Economics*, Vol. **17**(1), (2014).

Jörin, Robert and Lengwiler, Yvan. 'Learning from financial markets: auctioning tariff-rate quotas in agricultural trade', *Swiss Journal of Economics and Statistics*, Vol. **140**(4), (2004) pp. 521–541.

Khorana, Sangeeta. 'The developmental relevance of tariff rate quotas as a market access instrument: an analysis of Swiss agricultural imports', *Estey Centre for Law and Economics in International Trade*, Vol. **9**(2), (2008) pp. 1–24.

Lechner, Michael. 'The estimation of causal effects by difference-in-difference methods', *Foundations and Trends in Econometrics*, Vol. **4**(3), (2011) pp. 165–224.

Lence, Sergio H. and Hayes, Dermot J. 'U.S. farm policy and the volatility of commodity prices and farm revenues', *American Journal of Agricultural Economics*, Vol. **84**(2), (2002) pp. 335–351.

Logatcheva, Katja, van Galen, Michael, Janssens, Bas, Rau, Marie Luise, Baltussen, Willy, van Berkum, Siemen, Mann, Stefan, Ferjani, Ali, and Cerca, Mariana. 'Factors driving up prices along the food value chain in Switzerland: case studies on bread, yoghurt, and cured ham', *Strukturberichterstattung,* Vol. **60**(3), (2019).

Loi, Alberico, Esposti, Roberto, Gentile, Mario, Bruni, Mauro, Saguatti, Annachiara, Berisio, Serena, Cuppari, Luca, Aragrande, Maurizio. 'Policy evaluation of tariff rate quotas', Report mandated by the Swiss Federal Office of Agriculture, 2016.

Mann, Stefan and Lanz, Simon. 'Happy Tinbergen: Switzerland's new direct payment system', *EuroChoices*, Vol. **12**(3), (2013) pp. 24–28.

Márquez-Ramos, Laura and Martínez-Gómez, Victor D. 'On the effect of EU trade preferences', *New Medit*, Vol. **15**(2), (2016).

Martinez-Gomez, Victor, Garcia-Álvarez-Coque, José-Maria, and Villanueva, M. 'A trade model to evaluate the impact of trade liberalisation on EU tomato imports', *Spanish Journal of Agricultural Research*, Vol. **7**(2), (2009) pp. 235–247.

Moschini, Giancarlo. 'Economic issues in tariffication: an overview', *Agricultural Economics*, Vol. **5**(2), (1991) pp. 101–120.

OECD. *Agricultural Policy Monitoring and Evaluation 2016* (Paris: OECD Publishing, 2016).

OECD. *Agricultural Policy Monitoring and Evaluation 2017* (Paris: OECD Publishing, 2017).

OECD. *Agricultural Policy Monitoring and Evaluation 2019* (Paris: OECD Publishing, 2019a).

OECD. *Innovation, Productivity and Sustainability in Food and Agriculture* (Paris: OECD Publishing, 2019b).

OECD. *Agricultural Policy Monitoring and Evaluation 2020* (Paris; OECD Publishing, 2020).

Roth, Jonathan. 'Pre-test with caution: event-study estimates after testing for parallel trends', Working Paper, 2019.

Rubin, Donald B. 'Estimating causal effects of treatments in randomized and nonrandomized studies', *Journal of Educational Psychology*, Vol. **66**(5), (1974) pp. 688–701.

Santeramo, Fabio G. and Cioffi, Antonio. 'The entry price threshold in EU agriculture: deterrent or barrier?' *Journal of Policy Modeling*, Vol. **34**(5), (2012) pp. 691–704.

Schmitz, Troy G. 'Impact of the 2014 Suspension Agreement on sugar between the United States and Mexico', *Agricultural Economics*, Vol. **49**(1), (2018) pp. 55–69.

Skully, David W. *The Economics of TRQ Administration* (1999).

Skully, David W. *The Economics of TRQ Administration: ERS Technical Bulletin No. 1893, April* (2001).

Soon, Byung M. and Thompson, Wyatt. 'Nontariff measures and product differentiation: hormone-treated beef trade from the United States and Canada to the European Union', *Canadian Journal of Agricultural Economics/Revue canadienne d'agroeconomie*, Vol. **67**(4), (2019) pp. 363–377.

SWISSCOFEL. 'Leitfaden Importregelung Früchte und Gemüse 2019', 2018. http://www.swisscofel.ch/wAssets/docs/Dokumente-fuer-Website/2018/51-2018/Leitfaden-2019.pdf.

Thompson, A. Keith. *Fruit and Vegetables: Harvesting, Handling, and Storage,* 2nd ed. (Oxford, UK, Ames, Iowa: Iowa State Press, 2003).

Wright, Brian D. 'The economics of grain price volatility', *Applied Economic Perspectives and Policy*, Vol. **33**(1)**,** (2011) pp. 32–58.

WTO. *The WTO Agreements Series: Agriculture*. 3rd edition (2016).

Yang, Jian, Haigh, Michael S., and Leatham, David J. 'Agricultural liberalization policy and commodity price volatility: a GARCH application', *Applied Economics Letters*, Vol. **8**(9), (2001) pp. 593–598.

## VIII. APPENDIX

*Appendix 1. Seasonality in EU tariffs (incl. EU's EPS) for imports from third countries*

| Product | | EU system | | | | |
|---|---|---|---|---|---|---|
| Switzerland | European Union | Periods | Ad Valorem Rate | WTO Quota | Max. EP, euro per 100 kg | Min. EP, euro per 100 kg |
| • Cherry tomato<br>• Peretti tomato<br>• Tomato<br>• Truss tomato | Tomatoes, fresh or chilled | 1.1 - 31.3<br>1.4 - 30.4<br>1.5 - 14.5<br>15.5 - 31.5<br>1.6 - 31.9<br>1.10 - 31.10<br>1.11 - 20.12<br>21.12 - 31.12 | 0.088<br>0.088<br>0.088<br>0.14<br>0.14<br>0.14<br>0.088<br>0.088 | yes<br>yes<br>yes<br>yes<br>yes<br>yes<br>yes<br>yes | 77.8<br>103.6<br>66.8<br>66.8<br>48.4<br>57.6<br>57.6<br>62.2 | 84.6<br>112.6<br>72.6<br>72.6<br>52.6<br>62.6<br>62.6<br>67.6 |
| • Cauliflower, purple<br>• Romanesco<br>• Cauliflower<br>• Sprouting broccoli | Cabbage, cauliflower, kohlrabi, kale and similar edible brassicas, fresh or chilled | 1.1 - 14.4<br>15.4 - 30.11<br>1.12 - 31.12 | 0.096<br>0.136<br>0.096 | no<br>no<br>no | -<br>-<br>- | -<br>-<br>- |
| • Head lettuce | Lettuce and chicory, fresh or chilled | 1.1 - 31.3<br>1.4 - 30.11<br>1.12 - 31.12 | 0.104<br>0.12<br>0.104 | no<br>no<br>no | -<br>-<br>- | -<br>-<br>- |
| • Celeriac, soup<br>• Celeriac | Carrots, turnips, salad beetroot, salsify, celeriac, radishes and similar edible roots, fresh or chilled | 1.1 - 30.4<br>1.5 - 30.9<br>1.10 - 31.12 | 0.136<br>0.104<br>0.136 | no<br>no<br>no | -<br>-<br>- | -<br>-<br>- |
| • Greenhouse cucumber<br>• Slicing cucumber<br>• Cucumber, other | Cucumbers and gherkins, fresh or chilled | 1.1 - 31.2<br>1.3 - 30.4<br>1.5 - 15.5<br>16.5 - 30.9<br>1.10 - 31.10<br>1.11 - 10.11<br>11.11 – 31.12 | 0.128<br>0.128<br>0.128<br>0.16<br>0.16<br>0.128<br> | yes<br>yes<br>yes<br>no<br>no<br>yes<br> | 62.1<br>101.7<br>44.3<br>44.3<br>62.8<br>62.8<br>55.7 | 67.5<br>110.5<br>48.1<br>48.1<br>68.3<br>68.3<br>60.5 |
| • Snow pea<br>• Pea | Leguminous vegetables, shelled or unshelled, fresh or chilled | 1.1 - 31.5<br>1.6 - 31.8<br>1.9 - 31.12 | 0.08<br>0.136<br>0.08 | no<br>no<br>no | -<br>-<br>- | -<br>-<br>- |
| • Asparagus bean<br>• Bean extra-fine | Leguminous vegetables, fresh or chilled | 1.1 - 30.6<br>1.6 - 30.9<br>1.10 - 31.12 | 0.104<br>0.136<br>0.104 | no<br>no<br>no | -<br>-<br>- | -<br>-<br>- |
| • Artichoke | Other vegetables, fresh or chilled | 1.1 - 31.5<br>1.6 - 30.6<br>1.7 - 31.10<br>1.11 - 31.12 | 0.104<br>0.104<br>0.104<br>0.104 | no<br>no<br>no<br>no | 76<br>60.2<br>-<br>86.8 | 82.6<br>65.4<br>-<br>94.3 |
| • Courgettes /Summer squash | Other vegetables, fresh or chilled | 1.1 - 31.1<br>1.2 - 31.3<br>1.4 - 31.5<br>1.6 - 31.7<br>1.8 - 31.12 | 0.128<br>0.128<br>0.128<br>0.128<br>0.128 | no<br>no<br>no<br>no<br>no | 44.9<br>38<br>63.7<br>38<br>44.9 | 48.8<br>41.3<br>69.2<br>41.3<br>48.8 |

**Source:** Commission Implementing Regulation (EU) No 927/2012 of 9 October 2012 amending Annex I to Council Regulation (EEC) No 2658/87. Official Journal of the European Union (in force, access: 16 June 2020). Chapter 7.
**Note:** 'EP' stands for 'EU's entry price'. For the volumes imported by unit price lower than the minimum EP, the highest specific rate (for each 100 kg of imports) is applied. For the volumes imported by unit price higher than the maximum EP, the specific rate is zero. There are thresholds for prices between the minimum EP and the maximum EP where the specific rate gradually increases with the decrease of prices.

*Appendix 2. Switzerland's importance as an export destination for its neighbours*

The employed difference-in-differences approach requires that Switzerland's system of seasonal TRQs has no effects on vegetable producer prices in the comparison groups. Switzerland is considered a small country and, therefore, a price-taker in most international trade contexts. Ideally, we would like to assess the importance of Switzerland for production and price-setting decisions in the regions of Switzerland's neighbours used in the study. However, the necessary regional price, production and export statistics are largely unavailable. We therefore look at total exports, exports to Switzerland and domestic production at the national level. For this purpose, we draw on export figures from the UN Comtrade, FAO database and FAO production statistics. We calculated the share of exports of the three comparison countries, France, Germany and Italy, in their total production from 2014 to 2017 for six vegetables from our results section. The FAO statistics are not detailed enough to consider other vegetables. The results are shown in the graph in Figure A1. For 12 vegetables, which roughly correspond to the vegetables shown in our results section, the share of exports to Switzerland in the total exports of the three neighbouring countries from 2014 to 2019 can be calculated based on the UN Comtrade data (Figure A2).

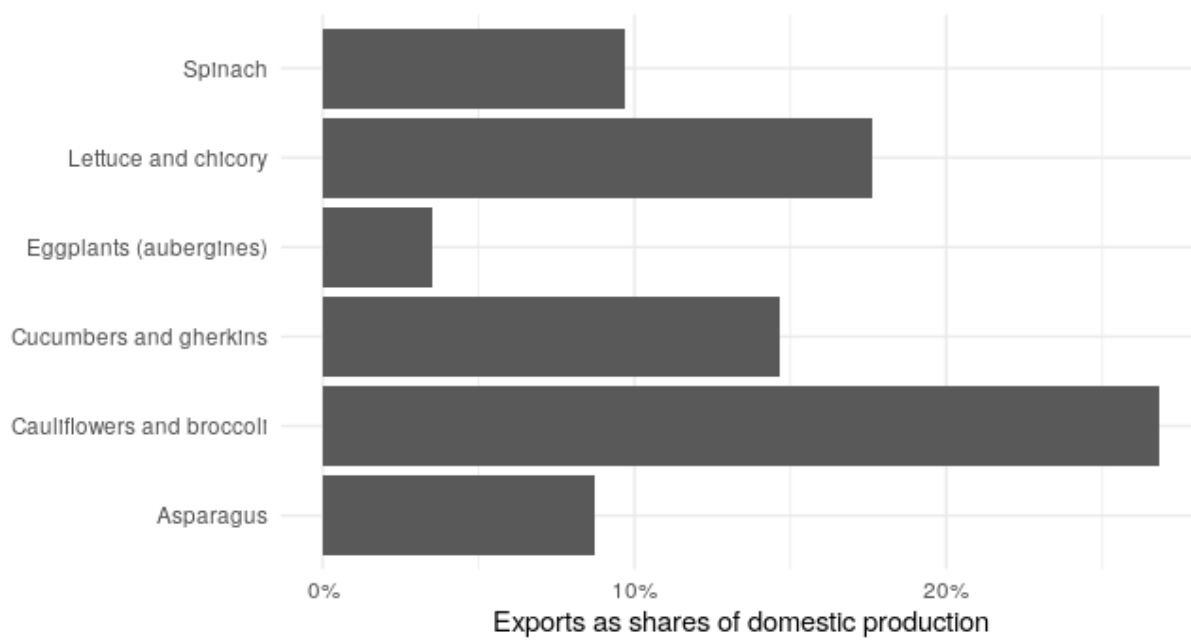

*Figure A1. Swiss exports as shares of total Swiss domestic production*

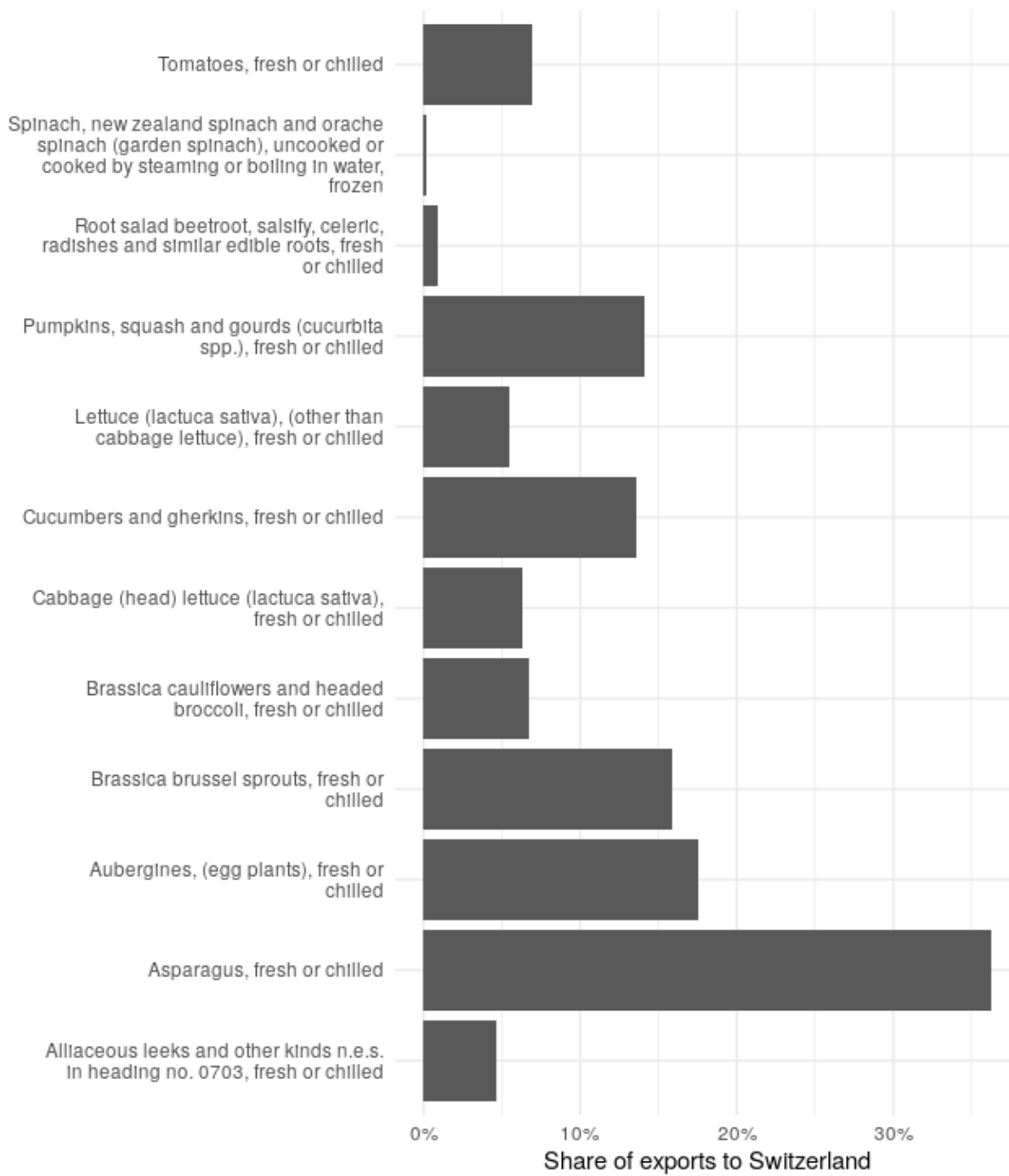

*Figure A2. Share of exports to Switzerland in total exports of France, Germany and Italy*

*Appendix 3. List of vegetables unsuitable for the estimation approach*

| Vegetable |
| --- |
| Beef tomato |
| Broad-leaved endive 'lavata' |
| Carrot (fresh) |
| Celeriac (fresh) |
| Courgettes yellow kg |
| Cucumber mini |
| Curled-leaved endive |
| Green celery |
| Head lettuce |
| Iceberg lettuce |
| Leek long stem |
| Lollo lettuce green |
| Lollo lettuce red |
| New onion |
| Peretti tomato |
| Sprouting broccoli |
| White and red onion |
| White cabbage (fresh) |

**Notes:** The difference-in-differences estimator requires a sufficient number of observations for both the treatment and the control groups, both within and outside the treated period. No difference-in-differences can therefore be determined for the vegetables listed.

*Appendix 4. Robustness tests*

We present in the text the results for the difference-in-differences estimator based on inverse probability weighting with seasonal and biweekly fixed effects for the average treatment effect of the protected phase. We performed additional robustness tests with other specifications, such as seasonal fixed effects only. In principle, our setting allows an estimation of the week-specific effects, that is, the effect of the TRQ system in weeks one, two and so forth of the protected phase individually compared with the last week before the start of the protected phase. We performed the estimation of the weekly effects; however, we do not include a graphical representation as it is space-consuming and did not generate significant insights. With a maximum of six observations per group (treatment and control) each week, it is not surprising that most effects are not significantly different from zero. In addition, no specific patterns for effects by weeks are observed. The corresponding results are available on request from the authors. Table A3 presents the results of the difference-in-differences estimation based on an OLS regression with seasonal and biweekly fixed effects. The estimated effects closely follow the ones estimated using inverse probability weighting.

*Table A3. Price effects of seasonal TRQs on vegetable prices estimated by OLS*

| | | Conventional | | Organic | |
|---|---|---|---|---|---|
| | | Level | Volatility | Level | Volatility |
| Vegetable | Comparison | Effect (standard error) | Effect (standard error) | Effect (standard error) | Effect (standard error) |
| Batavian lettuce | France | 19 (5.36)*** | 0.0101 (0.031) | -6.03 (5.67) | 0.0473 (0.0284)* |
| Beef tomato | France | 30.1 (6.76)*** | 0.045 (0.0685) | | |
| Broadleaf endive | France | | | 4.77 (11.2) | |
| Broadleaf endive | Germany | | | 2.85 (7.85) | |
| Brussels sprout | Germany | 7.29 (3.46)** | 0.049 (0.0101)*** | -4.53 (4.68) | 0.0588 (0.0101)*** |
| Carrots (fresh) | Germany | | | 15.3 (4.53)*** | -0.0106 (0.0434) |
| Carrots (fresh) | Italy | | | -18.1 (24) | -0.00776 (0.105) |
| Cauliflower | Germany | 23.1 (11.6)* | | 18.2 (11.6) | -0.0807 (0.125) |
| Cauliflower | Italy | -59.3 (27.8)** | | -32.6 (27.2) | |
| Celeriac (fresh) | Italy | -23.5 (Inf) | | -21.6 (3.4)** | |
| Cherry tomato | Germany | 47 (2.25)*** | 0.0279 (0.0132)** | 1.87 (1.87) | 0.064 (0.0122)*** |
| Cucumber (mini) | Germany | 14.1 (4.27)*** | 0.015 (0.0522) | | |
| Eggplant | Germany | 20.3 (4.8)*** | 0.0219 (0.0387) | -1.98 (4.84) | 0.0657 (0.0358)* |
| Eggplant | Italy | 20.7 (12)* | 0.107 (0.118) | 13.4 (11) | 0.114 (0.0988) |
| Florence fennel | Italy | | | -33.2 (18.4)* | |
| Greenhouse cucumber | France | 17.6 (4.36)*** | 0.0301 (0.0197) | -21.8 (5.11)*** | -0.00186 (0.0222) |
| Greenhouse cucumber | Germany | 19.8 (4.72)*** | 0.025 (0.0274) | -20.5 (5.46)*** | -0.0307 (0.0337) |
| Leeks | France | -4.72 (3.5) | 0.024 (0.00989)** | 0.00465 (3.29) | 0.0138 (0.00933) |
| Leeks | Germany | 2 (2.86) | 0.00286 (0.00991) | 1.76 (2.68) | -0.000314 (0.00937) |
| Leeks | Italy | -28.9 (4.51)*** | 0.05 (0.0252)** | -27.5 (5.35)*** | 0.0494 (0.029)* |
| New onion | Germany | -64.6 (24.5)** | | | |
| Oakleaf lettuce (green) | France | 25.4 (7.98)*** | 0.0649 (0.0495) | | |
| Oakleaf lettuce (red) | France | 18.9 (5.58)*** | 0.00173 (0.0315) | | |
| Radish (white) | France | 6.03 (8.17) | -0.00753 (0.0665) | 10.9 (5.25)** | |
| Slicing cucumber | Italy | -29 (15.4)* | 0.144 (0.178) | -17.3 (16.7) | 0.134 (0.156) |
| Spinach | France | | | 44.6 (5.1)*** | 0.0319 (0.0217) |
| Spinach | Italy | | | 5.39 (11.1) | |
| Tomato | Germany | 35.3 (4.21)*** | 0.0335 (0.0355) | 21 (5.28)*** | 0.0595 (0.0466) |
| Tomato | Italy | 91.2 (10.2)*** | -0.205 (0.0755)*** | 90.8 (13.3)*** | -0.0954 (0.0839) |
| Truss tomato | France | 29.5 (3.92)*** | -0.0498 (0.0293)* | -2.4 (5.12) | -0.0198 (0.0487) |
| Truss tomato | Germany | 39.9 (2.06)*** | 0.0101 (0.0124) | 14.7 (2.3)*** | 0.0192 (0.0159) |
| Courgettes /summer squash | Germany | 21.7 (10.7)** | 0.0431 (0.0986) | 2.49 (10.3) | 0.0981 (0.117) |
| Courgettes /summer squash | Italy | 42 (15.2)*** | 0.149 (0.0993) | -5.29 (13.9) | 0.0147 (0.105) |

**Notes:** Difference-in-differences estimates were obtained using OLS. The column 'effects' denotes the ATET for the protected phase either regarding the standardised price (a price of 100 corresponds to the average weekly price per season of the respective vegetable) or the relative weekly price change. Data for each vegetable were pooled by phases (protected/unprotected), and seasonal and biweekly fixed effects were employed. ***, **, * denote *p*-values smaller than .1, .05 and .01, respectively.

*Table A4. Price effects of seasonal TRQs on vegetable prices estimated with a propensity scores of being treated in the protected period of less than 0.99*

| | | Conventional | | | | Organic | | | |
|---|---|---|---|---|---|---|---|---|---|
| | | Level | | Volatility | | Level | | Volatility | |
| Vegetable | Comparison | Effect (standard error) | Pre-trend | Effect (standard error) | Pre-trend | Effect (standard error) | Pre-trend | Effect (standard error) | Pre-trend |
| Batavian lettuce | France | 19.7 (13.6) | | 0.007 (0.037) | | -6.29 (12.6) | | 0.046 (0.013)*** | |
| Broadleaf endive | France | | | | | 20.1 (15.02) | | | |
| Broadleaf endive | Germany | | | | | 3.01 (3.87) | | | |
| Brussels sprout | Germany | 7.54 (3.6)** | | 0.049 (0.023)** | | -7.29 (5.58) | | 0.059 (0.0116)*** | |
| Cauliflower | Germany | | | | | 60.7 (14.8)*** | | | |
| Cauliflower | Italy | | | | | 1.87 (23.7) | | | |
| Cherry tomato | Germany | 35.7 (6.86)*** | 16.6 (4.48)*** | 0.028 (0.016)* | -0.044 (0.044) | 10.6 (4.48)** | 8.84 (4.25)** | 0.064 (0.012)*** | |
| Eggplant | Germany | 25.6 (5.23)*** | | 0.021 (0.026) | | 28.1 (16.2)* | | 0.0642 (0.0326)** | |
| Eggplant | Italy | 29.6 (6.97)*** | | | | 29.3 (13.2)** | | | |
| Florence fennel | Italy | | | | | -22.2 (8.51)*** | | | |
| Greenhouse cucumber | France | 20.6 (3.04)*** | 8.86 (6.44) | 0.024 (0.0184) | -0.029 (0.029) | -18.9 (5.19)*** | -2.76 (6.97) | -0.006 (0.022) | 0.061 (0.038) |
| Greenhouse cucumber | Germany | 19.8 (3.61)*** | 9.75 (6.42) | 0.025 (0.027) | 0.0131 (0.035) | -20.5 (6.76)*** | -0.239 (7.61) | -0.031 (0.029)* | 0.12 (0.042)*** |
| Leek | France | 2.53 (11.5) | -8.84 (10.4) | 0.026 (0.012)** | -0.007 (0.017)* | 4.24 (10.7) | -10.6 (11.3) | 0.014 (0.014) | 0.010 (0.016) |
| Leek | Germany | 1.46 (10.7) | -0.652 (6.84) | 0.007 (0.013) | 0.002 (0.016) | 2.35 (9.7) | -0.454 (7.05) | 0.001 (0.01) | 0.028 (0.0125)** |
| Leek | Italy | -31.2 (18.01)* | 1.6 (10.3) | 0.054 (0.0170)*** | -0.043 (0.056) | -28.8 (18.8) | | 0.052 (0.009)*** | |
| Oakleaf lettuce (green) | France | 25.6 (15.7) | | | | | | | |
| Oakleaf lettuce (red) | France | 18.9 (12.5) | | -0.001 (0.01) | | | | | |
| Radish (white) | France | 35.7 (11.5)*** | | | | | | | |
| Slicing cucumber | Italy | -29.2 (19.5) | | | | -13.4 (15.6) | | | |
| Spinach | France | | | | | 14.1 (16.7) | | 0.029 (0.010)*** | |
| Spinach | Italy | | | | | 51.4 (10.87)*** | | | |
| Tomato | Germany | 36.2 (5.5)*** | | 0.046 (0.0347) | | 17.7 (4. 38)*** | | 0.058 (0.035) | |
| Tomato | Italy | 91.2 (1.4)*** | | -0.205 (0.01)*** | | 90.8 (7.17)*** | | -0.095 (0.0050)*** | |
| Truss tomato | France | 33.3 (6.23)*** | 28.7 (8.2)*** | -0.050 (0.025)** | -0.062 (0.047) | -2.4 (3.33) | | -0.020 (0.038) | |
| Truss tomato | Germany | 41 (1.27)*** | 11.9 (5.03)** | 0.010 (0.0125) | 0.025 (0.029) | 14.7 (5.22)*** | | 0.019 (0.015) | |
| Courgettes/summer squash | Germany | 2.1 (11.3) | | 0.015 (0.01462) | | -34 (21.1) | | 0.098 (0.061) | |
| Courgettes/summer squash | Italy | -5.07 (22.6) | | 0.149 (0.122) | | -33.8 (17.4)* | | 0.015 (0.020) | |

**Notes:** Difference-in-differences estimates were obtained using inverse probability weighting. The column 'effects' denotes the ATET for the protected phase either regarding the standardised price (a price of 100 corresponds to the average weekly price per season of the respective vegetable) or the relative weekly price change. For the main effect estimation, data for each vegetable were pooled by phases (protected/unprotected), and seasonal and biweekly fixed effects were employed. For the pre-trend test, observations from four and three weeks before the protection started were employed as the pre-period, while observations from the last two weeks before the protection started were used as a pseudo treatment period. No covariates were used in the pre-trend test estimation. ***, **, * denote *p*-values smaller than .01, .05 and .1, respectively. In the EU, greenhouse cucumbers and all types of tomatoes are protected by seasonal entry prices and seasonal tariffs, whereas cauliflower has only a seasonal ad valorem tariff and courgettes have only seasonal entry prices (see Appendix 1). Observations with propensity scores of being treated in the protected period that are larger than 0.99 are dropped from the sample to ensure common support.

*Appendix 5. Effect heterogeneity*

As evidenced by Table 1, seasonal TRQs affect domestic producer prices and levels. As a back-of-the-envelope calculation, we search for the main drivers of the effect heterogeneity. We acknowledge that the number of observations is low for such an endeavour. Nevertheless, we run a simple ordinary least squares (OLS) regression with the estimates of Table 1 as dependent variables.[23] As explanatory variables, we use a dummy variable *conventional* for quality, taking *organic* as a reference category, and we control for the comparison countries (variables *Germany* and *Italy*, with *France* as the reference category). In addition, we use a dummy variable that takes a value of 1 for vegetables harvested once a year. We hypothesise that a narrow harvesting window with one harvest increases production risks and potentially leads to a quick expansion of the offered quantities and, hence, increases pressure on the prices. For the same reason, we include storability measured in weeks, according to Thompson (2003). Moreover, we control for the length of the protected phase measured in days, and finally, we include the average market share of the vegetables and quality variety from 0 to 1 as niche products may be more resistant to price effects. The *market share* is defined for each vegetable of each quality as an average percentage of revenues from the vegetable in total revenues in the sample during the period from 2013 to 2019.

*Table A5. Determinants of the effect magnitude*

| Effect (standard error) | Price level | | | Volatility | | |
|---|---|---|---|---|---|---|
| | Pooled | Conventional | Organic | Pooled | Conventional | Organic |
| Intercept | -1.6218 (9.7942) | 29.8928* (14.9471) | 19.2225 (16.3837) | -0.0072 (0.0234) | 0.0289 (0.0487) | 0.0150 (0.0377) |
| Conventional | 16.8435* (8.4514) | | | 0.0141 (0.0207) | | |
| Germany | 10.4664 (7.3967) | -0.5961 (9.3908) | 9.7695 (10.9969) | 0.0175 (0.0161) | -0.0018 (0.0277) | 0.0262 (0.0200) |
| Italy | 11.7764 (8.3296) | -4.5935 (11.8557) | 3.9691 (12.8743) | -0.0269 (0.0227) | -0.0384 (0.0414) | -0.0326 (0.0290) |
| Harvested once | 15.5182* (8.6849) | 1.3245 (13.7718) | -6.9200 (14.9091) | 0.0087 (0.0222) | -0.0381 (0.0421) | 0.0141 (0.0358) |
| Storability | -7.0271* (3.7335) | -3.9025 (5.1977) | -6.3924 (5.2720) | 6.3e-04 (0.0086) | 0.0085 (0.0155) | -0.0034 (0.0096) |
| Market share | 1.92 (3.11) | 1.41 (3.09) | -41.12** (17.13) | -0.0038 (0.0071) | -0.0089 (0.0097) | -0.03 (0.03) |
| Days protection | -0.0426 (0.0548) | -0.0757 (0.0752) | 0.0460 (0.0812) | 1.2e-04 (1.3e-04) | 1.8e-04 (2.4e-04) | 9.4e-05 (1.4e-04) |
| n. Obs. | 72 | 33 | 39 | 54 | 26 | 28 |
| R2 | 0.2092 | 0.1699 | 0.3043 | 0.1193 | 0.1102 | 0.3849 |

**Notes:** The OLS regressions employed the estimated coefficients from Table 1 as dependent variables. ***, **, * denote *p*-values smaller than .01, .05 and .1, respectively.

---

[23] Note again that we use results from first-stage estimations as dependent variables and test several hypotheses simultaneously (see e.g. Bonferroni, 1936). The reported OLS standard errors should, therefore, be treated with caution.

In Table A5, we present the results explaining the TRQ effects on price level and volatility, respectively. Subsample estimations for organic and conventional production must be interpreted with caution due to the very small number of observations. The estimates confirm that the seasonal TRQs have a stronger impact of roughly 16 percentage points on the price support for conventional production. The more perishable the vegetable is, the more it profits from the price support of TRQs because an additional week of storability decreases the effect by seven percentage points. The analysis does not suggest that niche vegetables differ systematically from mass products or that the length of the protected phase alters the effect of seasonal TRQs. While the subsample estimates tend to show some effect heterogeneity, the differences in effects between conventional and organic products are not significant. The coefficient for market share is significant for prices of organic products. Note the market shares of organic vegetables considered in Table A5 range from 0.02 to 1.1 %. The coefficient of -41.12 implies that the price increasing effect of the protected period is 19.1 standardised price units (which can be interpreted as 19.1 % of the average weekly producer price) lower for an organic vegetable with a high market share (defined as the third quartile of the market share variable for organic vegetables) than for an organic vegetable with an average market share. Except for the country effects, our back-of-the-envelope analysis cannot uncover the determinants that explain for which vegetables the effects of TRQs on the week-to-week price volatility are stronger.